# Semi-Automated & Collaborative Online Training Module For Improving Communication Skills

Ru Zhao, Vivian Li, Hugo Barbosa, Gourab Ghoshal, Mohammed (Ehsan) Hoque, UNIVERSITY OF ROCHESTER

This paper presents a description and evaluation of the ROC Speak system, a platform that allows ubiquitous access to communication skills training. ROC Speak (available at rocspeak.com) enables anyone to go to a website, record a video, and receive feedback on smile intensity, body movement, volume modulation, filler word usage, unique word usage, word cloud of the spoken words, in addition to overall assessment and subjective comments by peers. Peer comments are automatically ranked and sorted for usefulness and sentiment (i.e., positive vs. negative). We evaluated the system with a diverse group of 56 online participants for a 10-day period. Participants submitted responses to career oriented prompts every other day. The participants were randomly split into two groups: 1) treatment - full feedback from the ROC Speak system; 2) control – written feedback from online peers. When judged by peers (p<.001) and independent raters (p<.05), participants from the treatment group demonstrated statistically significant improvement in overall speaking skills rating while the control group did not. Furthermore, in terms of speaking attributes, treatment group showed an improvement in friendliness (p<. 001), vocal variety (p<.05) and articulation (p<.01).

## 1. INTRODUCTION

From professional settings to casual interactions, speaking skills are vital in everyday life. Being able to effectively communicate ideas and thoughts are considered important skills that are valued in almost every profession. Given the importance of communication skills, many entities exist such as Toastmasters [1] and

Author's address: Ru Zhao, email: rzhao2@u.rochester.edu; Vivian Li, vli4@u.rochester.edu; Hugo Barbosa, email: hbarbosa@pas.rochester.edu; Gourab Ghoshal, email: gghoshal@pas.rochester.edu, Mohammed(Ehsan) Hoque, mehoque@cs.rochester.edu, University of Rochester, Rochester, NY 14611t owntheroom! [2] that are intended to help individuals. While very effective, this approach eliminates a significant part of the population who are not comfortable seeking help yet love to practice and improve their communication skills. An automated communication skills training system that is available whenever they want, wherever they want would be invaluable to them.

How difficult it is to develop a system that can understand and recognize the behavioral nuances and provide respectful feedback? In verbal communication, emphasis is placed on the meaning of what is said, organization of ideas, and clarity. While these factors are vital, how information is presented through nonverbal behaviors (facial expressions, volume, gestures, eye contact) can reinforce or undermine a speaker's intentions. For example, speaking too fast can limit speech comprehension and indicate nervousness, however speaking too slowly affects the liveliness of a speech and lose the audience's attention [3]. Experts suggest that movement during a speech can make a speech more expressive or reinforce a point [4]. However, depending on the type of movement, it can also be a distraction [3, 4]. Pausing during a speech can indicate a speaker's careful planning of thoughts, but can also expose an unprepared speaker [7]. Varying pitch and volume can help clarify messages, pose questions, or establish the idea tone [3, 8]. Vocal cues can also express negative traits. Vocalized pauses or fillers such as "ah" or "um" are distractions that may express nervousness or lack of sincerity of the speaker [9]. The meaning of behavior could be strengthened or contradicted when they are combined with multimodality [10, 11]. For example, expressions of amusement can be detected by observing a speaker's facial expression but the effect can be bolstered by the speaker's voice [11]. Automated understanding, recognition and interpretation of these behaviors are still an open problem because we simply do not have the example data set to train automated systems that understand the cultural, behavioral and contextual nuances of human behavior.

In our previous work [12], we developed an automated system called ROC Speak which allows anyone with access to a computer visit a website, enable their webcam, practice communication skills and obtain feedback in a repeatable, standardized and respectful manner. While computer algorithms are more reliable at consistently and objectively sensing subtle human behavior, human intelligence is superior at interpreting contextual behavior. Therefore, in ROC Speak, computer algorithms perform the sensing portion while outsourcing the interpretation process to humans. This allowed us to develop a semi-automated behavior analysis system. Our system can automatically record and analyze videos for certain behavioral metrics (e.g., smiles, body movement, intonation spoken words etc.). In addition, users can share their videos with anyone that they trust (e.g., close friends, social network, online workers) for subjective interpretation of their nonverbal behavior. Our system automatically prioritizes the users' comments and presents users with those that are most helpful and constructive.

A comprehensive evaluation of ROC Speak or any other system that claims to improve human behavior introduces a set of challenges. First, it is important to recruit a significant number of participants and have them continue to use the system to understand the temporal trajectory of improvement over an extended period. This also could potentially tease out the "novelty factor" if people were to use the system once or twice. However, recruiting and retaining a significant number of participants is challenging. Another consideration is to deploy the intervention online so that participants could try it out in their environment with minimal interaction and support from the researchers. This approach ensures that if the intervention is successful, a real-world deployment may likely be successful as well.

In this paper, in addition to describing the technical pieces of ROC Speak, we provide an evaluation of the ROC Speak system in the context of job interview related prompts. We hired 56 online workers and randomly split them in two groups; 1) treatment; 2) control. Participants from both groups had to upload 5 videos in 10 days on job interview questions. We collaborated with counselors from the career services of the Rochester Institute of Technology to decide on the prompts. The participants in the treatment group received automated feedback from ROC Speak as well as provided anonymous comments to other participants which were automatically ranked and sorted for usefulness and sentiment (e.g., positive vs. negative). In the control group, participants reviewed each other videos and wrote anonymous comments for other participants. Our system allowed participants to record themselves giving a speech using their computer in their home environment. Researchers had no physical interaction with the users. All communication was conducted via email reminders for uploading videos and troubleshooting for technical issues. Evaluations were conducted using two metrics. First, participants anonymously rated each other after every prompt in both conditions (also known as peer ratings). Second, we hired an expert who labeled all the videos in both conditions in a random order (also known as independent judge). Both the peer and the independent judge's ratings confirmed that participants in the treatment group demonstrated statistically significant improvement in overall ratings in their speaking skills performance.

In particular, this paper makes the following contributions:

- An online intelligent interface that provides automated feedback to users and allows users to be a part of a community where they can practice, receive, and provide feedback to each other on their communication skills.
- The results from a 10-day study showing the effects of repeated training and the effectiveness of combined automated and subjective feedback.

## 2. Related Work

Many efforts have been made to create tools for improving social or presentation skills. We find Cicero [13], MACH [14], Rhema [15], ROC Speak [12], Automanner [16] and Automated Social Skills Trainer [17], Presentation Trainer [18], Logue [19], Tardis [20–22] to be the most recent examples. For example, Cicero [13, 23] simulates a virtual audience as part of live feedback to a speaker, Automanner [16] is an intelligent interface that makes users more aware of their body language by automatically extracting and displaying repeated patterns. MACH includes a reactive 3D character allowing anyone to practice job interviews and receive visual feedback at the end. While these systems adopt innovative ways to help users improve their communication skills, the evaluation of these systems does not reflect real world usage. We argue that, to create an unequivocally ubiquitous tool for improving communication skills, it is essential to consider the following aspects during evaluation.

Naturalistic Environment: While lab experiments are valuable in providing insight on specific aspects, bringing participants into the lab, asking them to perform a task, and have them self-reflect on the experience become unavoidable confounding factors. Therefore, if possible, it is desirable to conduct an experiment that is outside of the lab so that the results can generalize to real world usage [24]. The environment should be reflective of the one participants will be using the tool in.

*Diverse Users:* Most evaluation of the tools invites college student as participants from a university campus [14–16]. It is unknown whether the effect can be generalized to the general population.

*Robustness:* When an experiment is conducted in a lab setting, we often control for devices, environment and many other factors. While controlled experiments are useful to understand an exact effect [13–15, 17, 23] , it is also important to deploy the technology in the wild and make it work across devices, sensors, and environment. It allows us to understand the real-world constraints and design the technology accordingly.

*Multiple Usage:* Many technologies are subject to the novelty effect where users tend to temporarily improve due to the introduction of new technology instead of the actual response to the technology. In some cases, evaluation is done in one session with very little downtime. As a result, it becomes difficult to assess if the effect and interest will remain after. Multiple usage with appropriate down time in between can test against these factors.

*External Evaluation:* Self-reported data can be unreliable, especially when communication skills are being evaluated. For example, people with social anxiety generally rate their own performance being worse than when others evaluate them [25][26].

**Table I. Summary of the Evaluation Procedures by Communication Tools**

| | Naturalistic Environment | Diverse Participants | Robust-ness | Repetitive Usage | External Evaluation |
|---|---|---|---|---|---|
| Cicero [13] | | x | | * | x |
| Rhema [15] | | | | * | x |
| Automated Skills Trainer [17] | | | | | |
| Automanner [16] | | | | * | x |
| MACH [14] | | | | x | x |
| Presentation Trainer [18] | | | | * | x |
| Logue [19] | x | | | | x |
| Tardis [20–22] | | | | | x |
| ROC Speak [12] | x | x | x | | |
| ROC Speak (Current Evaluation) | x | x | x | x | x |

***The tool was repeatedly used, but all done in one long session with very short downtime.**

An examination of the current state of the art systems and the evaluation procedures of those systems, are summarized in table 1. To satisfy the repetitive usage criterion, users must have used the tool more than twice.

Most of these systems are very innovative. However, they are computationally intensive, sensitive to environmental factors, or requires devices (Microsoft Kinect [16, 18, 20] and Google Glass [15, 19]) that most people don't have access to. These limitations prevent them from being evaluated in a more realistic world setting. ROC Speak is a lightweight platform that is available for anyone that has access to a Chrome browser and a webcam. In our previous evaluation of ROC Speak [12], we were able to address some of these evaluation issues by deploying our system online and allowing users to practice our tool in their own environment. In this paper, we deployed our system online for repetitive usage by employing a training regime for a 10-day period and have peers and a judge to evaluate their performance. To our knowledge, this may be the first deployment of a communication skill training system in a real-world setting that has proved to be effective in improving a users' communication skills.

## 3. Design and implementation

In this section, we describe the design decisions and implementation details for our feedback interface. Our feedback system consists of two key components: subjective feedback generation and automated feedback generation. In our system, we encourage subjective peer feedback exchange by placing mechanics for comment anonymity and feedback minimum requirements. Meanwhile, the automated feedback immediately extracts audio and visual information from user videos and presents it to the user in a variety of visuals.

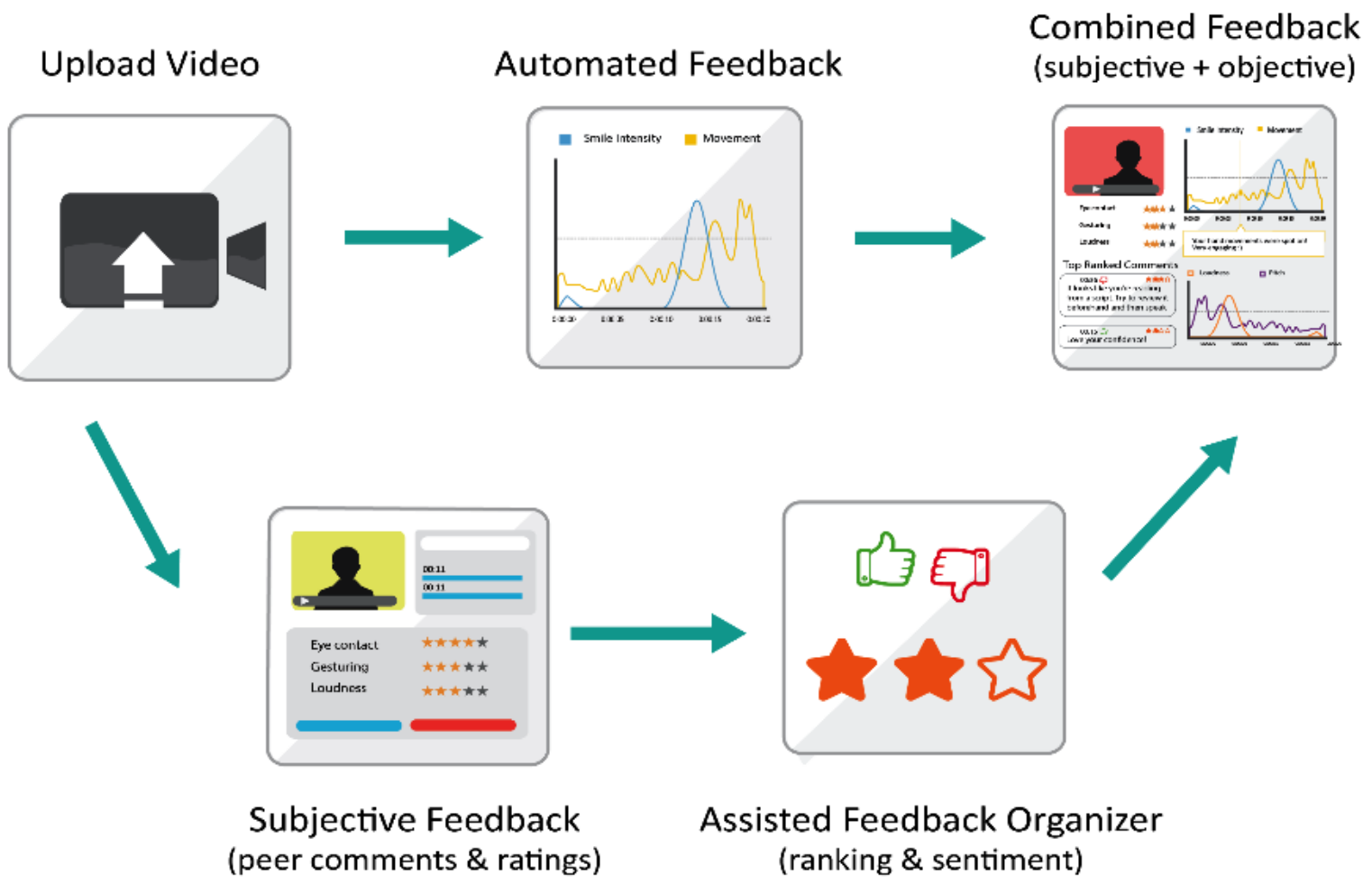

**Figure 1. The semi-automated feedback system immediately generates automated feedback after the video upload. Subjective feedback (ratings and comments) are gathered from peers and organized by our assisted feedback organizer. Lastly, automated and subjective feedback is combined.**

The overall workflow of our feedback interface generation is shown in Figure 1. After users upload a video in response to a prompt, they receive immediate automated feedback that is generated from the audio and visual information collected from the video. This feedback includes smile intensity, body movement, loudness and pitch, speaking rate, transcript, unique word count and word cloud, and weak word usage. All of this feedback is presented through a graph or visual that users can interact with to gain in-depth information. Once users receive a feedback (three comments and ratings) from their peers, our classifier labels the comments on usefulness and sentiment and adds them to the feedback interface. The most useful comments are prioritized at the top of the comment section. If participants receive a new peer feedback, they receive a notification.

### 3.1 Challenges and Motivations of Subjective Feedback Generation

While automated features are generated in a consistent manner, quality of subjective feedback can vary due to reviewer motivation, attention, and other confounding factors. This highlights the need for a quality assurance mechanism. It motivated our development for a community bonding experience that encourages quality feedback exchange. Additionally, participation in learning communities are positively related to enhanced performance and self-reported gains [27, 28]. These joint groups produce a culture of productive conversation and a shared sense of purpose from which participants benefit [29]. To promote the development of a learning community, we designed a social system that allows participants to engage with their peers and freely exchange opinions and commentary.

#### 3.1.1 Giving and Receiving Subjective Feedback

Giving and receiving feedback are the foundations for social interaction in our application. We allow participants to provide feedback to each other in two formats: 1) Comments as opinions written as short sentences or paragraphs; 2) Ratings on a scale of one to five stars.

When users leave comments for other participants, the comments are attributed to general presentation qualities (movement, friendliness, and speech) to help users recognize areas for improvement. Comments associated with a timestamp that reference the relevant point in the video has been shown to improve engagement with content [30], which we implemented in our system.

To make the ratings specific to each person, when a video is uploaded for peer review, the owner can select five presentation qualities, such as eye contact or pacing, out of a list of 23 qualities to be rated on by peers. This customization allows trainees to personalize their feedback, encouraging them to reflect on their speech and target areas for improvement. Also, if the user wants feedback that is not captured by the customized rating form, they can make an appeal in the video description, which is displayed in the video feed.

#### 3.1.2 Encouraging Interactions

To motivate users and encourage feedback contributions, we introduced the following design choices to the system:

- Anonymity is enforced in peer feedback. Video owners can see a pool of feedback; however, they are unable to connect ratings and comments with other users. We enforce this anonymity because human subject considerations required that we protect the identity of the users and prevent any potential conflict. This also encourages users to freely leave praise, critiques, and suggestions without the fear of generating disputes or judgment.
- We require a minimum of three comments for each video reviewed. This encourages users to pay attention to the videos and leave varied feedback.
- We require users to review (comment and rate) a minimum of three videos, each video represented as a point on the interface. If the minimum requirement is not met, users are unable to upload additional videos. This ensures user engagement with the community and peer feedback on each video upload.
- We implement a video feed page, displaying all unwatched video uploads by peers. This page is set as the default homepage to promote giving feedback and to allow users to see their peers' progress.

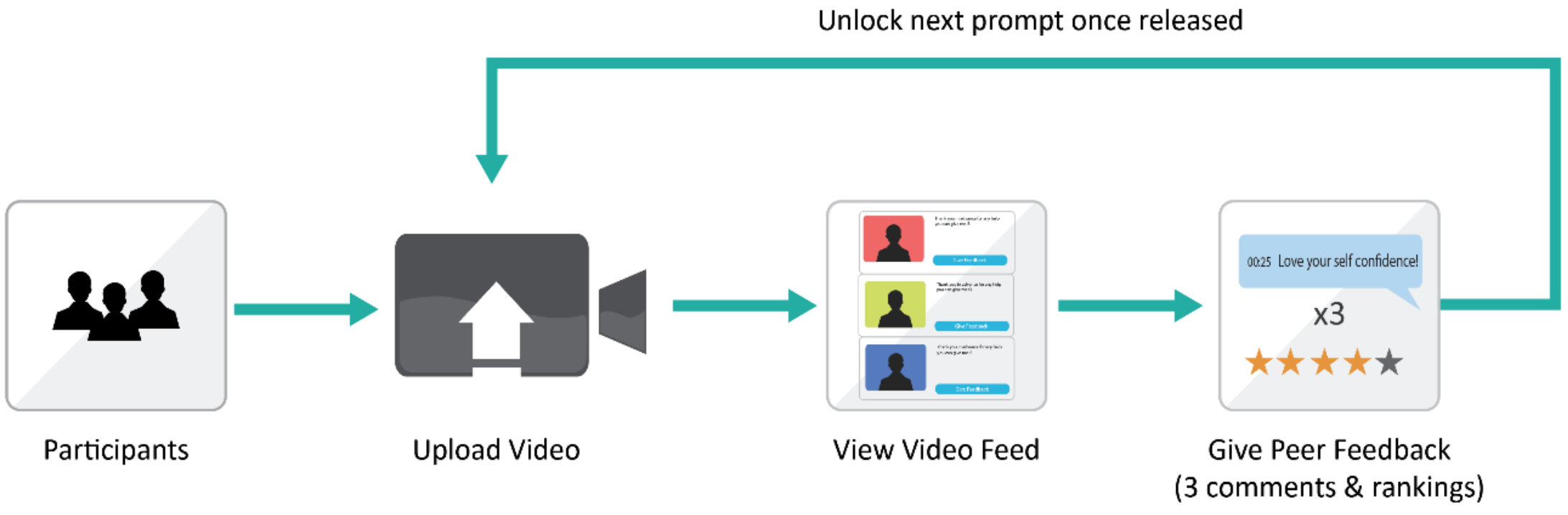

**Figure 2. Description of the study design and the workflow of our framework. Users must (1) record a video to be uploaded to the video feed, (2) provide feedback to three other users' videos and (3) wait until the release of next prompt to fully unlock it.**

Our system design choices are shown in Figure 2. Users begin by responding to a prompt and recording a video of their speech presentation. Users can then customize their videos with a title, description, and five presentation qualities they want to receive ratings on. Uploaded videos are displayed on a video feed where peers can comment on and rank the videos. After a user gives a minimum of three reviews, consisting of 3 comments and rankings, to different videos, they are able to upload a video for the following prompt. This design was integral to our study procedure. Further details on prompts and flow will be discussed in conditions and tasks.

## 3.2 The Feedback Interface

The feedback interface (Figure 3) allows users to review their recorded video alongside a combination of automated feedback and peer feedback. The automated feedback was generated using ROC Speak [12], which includes feedback on smile intensity, body movement, loudness, pitch, and word prosody. In addition to this feedback, we have added features for word cloud, unique word count, and total word count. Below, we describe what feedback is provided and the significance of each feedback as it relates to communication skills.

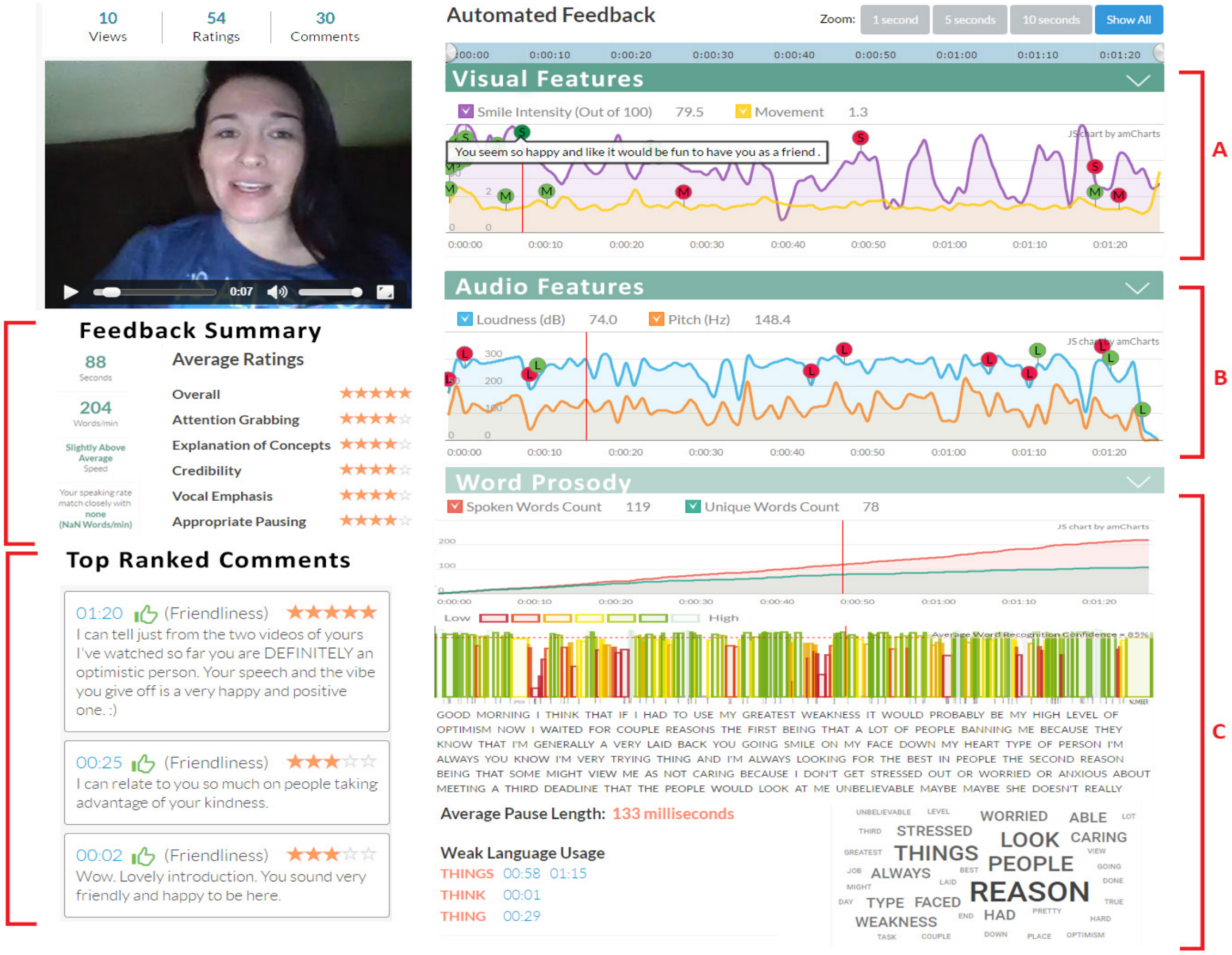

**Figure 3. Our feedback page allows users to review their speech and read a variety of feedback. The page shows automatically generated measurements for (A) smile intensity, gestural movements, (B) loudness, pitch, (C) unique word count, word recognition confidence, a transcription of the speech, word cloud, and instances of weak language. Additionally, we create (D) a Feedback Summary section, and (E) a top ranked comments section classified by usefulness and sentiment**

### 3.2.1 Smile Intensity

Smiling is integral in establishing trust and friendliness with an audience [31]. To help users gauge their smiling, we display measures for smile intensity in a line graph (Figure 3A). These measures are calculated by using Shore [32] facial recognition software to detect facial expression in each frame. Shore uses the Adaboost algorithm to recognize "smiling" or "neutral" using sample images. The outcome of the classifier is then normalized to generate a score that ranges from 0-100 (0 means no smile, 100 means a wide smile). More details on the extraction and evaluation of this feature are explained in previous work [12, 14].

### 3.2.2 Body Movement

Meaningful body movements can make a speech more engaging or help reinforce a point [4]. However, there are also unintentional body movements that can detract from a speech [5, 6]. We bring these movements to the user's attention by displaying measurements for body movements (Figure 3B). These measurements are computed using a background-subtraction technique that calculates pixel-wise differences between every adjacent pair of video frames. This technique works relatively well with the assumption that the speaker has a still background and the distance between the user and the computer is constant. From a previous study of ROC Speak, users have self-reported the measurement to be accurate and helpful [12]. While using background subtraction to describe movement was useful in our study, incorporating other information such as head nodding and hand waving remains part of our future work.

#### 3.2.3 Loudness and Pitch

How well a speaker projects their voice and emphasizes key information is crucial to presenting a confident and engaging speech [33]. For instance, speaking in monotone makes a person sound robotic whereas variations in pitch can show a broad range of emotions. To help make users aware of loudness and pitch, we present measurements in a line graph (Figure 3B). Measurements for loudness are extracted in decibels (dB) and pitch in Hz using *Praat* [34], an open source software for the analysis of speech. The accuracy of the pitch and volume are subject to noise in the user's environment and the quality of captured audio.

#### 3.2.4 Transcript and Word Prosody

The interface displays a transcript of their speech and a word recognition confidence value for each word (Figure 3C). Our framework uses the Google Web Speech API [35] and Speech Recognition SDK [36] to generate the transcript, and uses Nuance for word recognition confidence. The word recognition confidence displays how well the computer recognizes words. We also use the Penn Phonetics Lab Forced Aligner (P2FA) [37] to match the words of the transcript with the sounds in the recorded audio. The forced alignment provides the exact start and end times of each transcribed word. These times are used to find the average loudness and duration of each spoken word. There is no ground truth for proper articulation, but looking at these graphics users may get a sense what they want to improve. The accuracy of the transcript can affect the quality of these measurements.

#### 3.2.5 Unique Words and Word Cloud

Data for unique words and word cloud are extracted from the transcript and displayed on the interface (Figure 3C). Unique word count is a measure of number of unique words over total words and reveals how varied the vocabulary is. The word cloud highlights the most common words identified in the speech, often reflecting the theme of the speech. We count the most frequent words and display the results using a D3.js word cloud graph [38].

#### 3.2.6 Weak Word Usage

Using filler words during a speech is sometimes an indication of lack of preparation or nervousness and can be distracting [9]. To promote flowing speech, we display each filler word and a time stamp for each instance of the filler word (Figure 3C). We classified weak word usage using a set of acquired words from owntheroom!, a company that provides service to improve public speaking [39].

#### 3.2.7 Peer Feedback Usefulness and Sentiment Classification

A common issue with open forums is a stream of low-quality comments such as "That was good" that distract users from prioritizing more thoughtful and helpful feedback [40]. To address this issue, current online communities use methods such as user flagging and human moderation to filter comments. Further research in comment moderation suggests filtering for usefulness and sentiment to enhance user engagement [41, 42]. For that reason, we ranked peer comments in order of usefulness and placed the most relevant comments at the top (Figure 3E). In addition, we ranked the comments based on sentiment (positive vs. negative) with priority given to positive comment. To do this, we created an automated comment moderation system that immediately moderates comments as they are entered and curtails the need for a human moderator. This moderation system is composed of two classifiers: a helpfulness scorer and sentiment classifier.

*Data collection*: Data was collected by recruiting 25 Amazon Mechanical Turkers to record a speech. Then we recruited additional Turkers to watch and leave comments on the videos, resulting in at least 30 comments on friendliness, body language, and voice modulation for each video.

*Ground truth labels for Helpfulness Scorer:* Each comment was rated by 10 different Turkers with a number ranging from 1 to 4, with a score of 4 being the most helpful. For the ground truth label, we simply sum the ratings from the crowd to obtain a helpfulness score. For instance, a highly-ranked comment could be "You don't use many hand gestures, and you look off the screen often. Focus on the camera." (An unhelpful comment: "Good speech.") For our training and testing dataset, we obtained a total of 1,649 time-stamped comments with corresponding helpfulness scores. Each comment was ranked by 10 different Turkers on usefulness on a scale of 1 to 4.  For our training and testing dataset, we obtained a total of 1,649 time-stamped comments with corresponding helpfulness scores.

*Ground truth labels for Sentiment Classifier:* As the sentiment classifier aims to provide insights about helpful criticisms and praises, we excluded comments that were labeled most unhelpful—that is, those in the lowest quartile of the helpfulness ratings. A research assistant manually labeled the remaining comments as either positive or negative. Thus, we obtained a dataset with 742 positive comments and 285 negative comments.

*Ranking helpful comments:* We extracted two types of features: text and video. Based on observations, we noticed that helpful comments tend to be long, properly punctuated, and appropriately capitalized. As a result, our text features include: total number of characters, presence of punctuation in the comment, and presence of capital letters. We also extracted parts of speech information from the comment. In addition to extracting text information, we also integrate audio and visual information from the recorded videos. Multimodal signals have been very successful for predicting the quality and success of a speech [43][44], we used this intuition to predict the relevance of our comments. If a user smiles or moves more/less at a given time and if there are comments that appear at the location to address this change, such comments will be more likely to be relevant and helpful. Our labeling interface allows Turkers to indicate the time for a corresponding video. Therefore, given that a comment, the algorithm extracts features related to smile, loudness, movement and pitch around one, two and four second windows within the timestamp. We used a linear regression to predict the helpfulness score and report a coefficient of determination (R2) of .3 for body movement, .18 for volume and .24 for friendliness, where the coefficients correspond to the amount of the variance in the helpfulness accounted by each of the variables.

*Sentiment classification:* To recognize whether a comment is positive or negative, we extract unigram and bigrams as features for sentiment classification. Each comment is then transformed to a feature vector it its tf-idf representation. We applied a Naïve Bayes classifier over the training data to predict the sentiment of the comments. When 70% of all the data was used to train the sentiment classifier and 30% for testing, the classifier achieved an accuracy of 82.03%.

#### 3.2.8 Embedded Peer Feedback

Graphs for visual and audio features (Figure 3A) are annotated with markers at the original time a comment was made. When users hover over the marker, the comment appears.

## 4. Evaluating the Training System

In this section, we describe our study design decisions and assessment methods. Our system is designed to provide a comprehensive and collaborative training experience on job interviews comprised of a community environment, verbal and nonverbal evaluations, and assisted feedback organization. To test the effectiveness of our system, we sought to answer the following questions:

1. How valuable is our treatment (combined automated and subjective feedback) interface compared to the control (subjective feedback only) interface?
2. Is the system effective in facilitating the generation of subjective feedback and are those feedbacks useful?
3. How useful do people find the system?

### 4.1 Conditions and Tasks

To determine the effectiveness of our interface, we set up a user study in which participants were randomly divided into either the treatment conditions with all features (verbal and nonverbal measurements, assisted feedback organization, and feedback summaries) or the control condition that only shows peer comments and ratings and video playback. Thus, users could view and provide feedback to peer videos in both conditions but varied between the set of feedback that participants received. The control condition was intended to model the combined capability of commercially available products ratespeeches.com [45] and Vialogues [46], where users can exchange feedback on videos they share. However, both products lack an automated system to provide feedback or a process to maintain the quality of subjective feedback. Also, they are general purpose systems, unlike our system which specifically targets the purpose of improving communication skills.

To simulate the intended use of our system in speech training, we created several training tasks for users to complete when using the test interface. Each task consisted of two parts:

- Uploading a video via webcam in response to a prompt. Prior to uploads, participants watched a one-minute video that introduced guidelines to answering the prompt. To ensure reasonable familiarity and interest in the prompts, we adopted the most common interview questions, which were provided to us by the Career Center at the Rochester Institute of Technology (RIT). Additionally, there was no limitation to the number of uploads per prompt, allowing users to practice multiple times for a prompt. Only the final video of each prompt is taken into consideration for evaluation. The prompts, in order, are as follows:
    1. Tell me about yourself
    2. Describe your biggest weakness
    3. Tell me about your greatest achievement
    4. Describe a conflict or challenge you face
    5. Tell me about yourself

We require Prompt 1 and 5 to be the same in order to acquire a measurement for pre and post training.

- Giving feedback to a minimum of three peer videos where a feedback consists of a minimum of three comments and ratings.

These tasks were spread out over an extended duration of time to give participants repeated practice. However, to move on to the next task, participants had to complete the previous task.

### 4.2 Participants and Procedure

Fifty-six participants (27 females, 29 males, with ages between 18 to 54 years) were recruited from Amazon Mechanical Turk and paid $25 upon completion of the study. Originally, we had 67 participants but nine dropped out in the first few days of the study with three citing family or health reasons and the remaining 6 giving no reasons. Participants were screened through an online qualification video where they were screened for technical quality, as a webcam and microphone were necessary to complete the study. Participants came from a variety of educational backgrounds, ranging from high school graduates to postgraduate degree holders. A majority had some college experience. Prior to the study, 92% of participants reported that they practice public speaking less than 10 times a year, and 44% of the participants never practice.

Participants were randomly assigned to either the treatment or control condition. There were 15 females, 11 males in the control condition and 14 females and 16 males in the treatment condition. In both conditions, we introduced the study as an online interview preparation workshop that could be completed remotely. Users were given introductory instructions to the site and allowed to freely explore the interface. We conducted a 10-day study that included five prompts, which were released every two days. Although we allowed users to interact with the interface at any time, we set a deadline for each task every two days to correspond with the release of the next prompt.

We encouraged participant engagement with an incentive system. Participants could receive additional cash awards by being a top-10 performer ($20) and a top-10 commenter ($15). Participants were told that top performers would be determined through expert evaluation and reflected on a public leaderboard updated every two days. This setup encouraged participants to prepare for their videos and utilize the feedback page to improve their videos. Top commenters were outlined as being consistently active throughout the study, giving appropriate feedback, and giving feedback often. This award promoted engagement with peer videos and quality comments beyond the minimum required feedback.

## 4.3 Measures

### 4.3.1 User Assessment

Users in each group were asked to review videos from other users and rate the videos on overall performance. They were aware told that the reviews would be used as feedback to other users and were unaware that the ratings would be used for any other purposes. Users were also aware that reviews given to other users would remain anonymous (ratings had no user identifier).

### 4.3.2 Assessment by Independent Judge

To eliminate bias from having partial knowledge of our system, we hired a 3rd-year research assistant from the psychology department to independently evaluate all the videos. The videos from both groups were mixed together and presented to the evaluator in random order. The evaluator was blind to the context of the study. In addition to evaluating the overall performance of the person in the video, the independent evaluator was asked to rate aspects of the speech that are commonly found for indicators of a good speaker [47]. The survey can be found here - http://tinyurl.com/zf9nlpu.

### 4.3.3 Self-Assessment

Participants from both groups were asked to fill out a total of six surveys throughout the study: one initial, four after each completed prompt, and one final survey. These surveys included numerous 7-point Likert Scale and free-response questions about their perception of and experience with the system. In the initial survey, participants were asked about to provide demographic information and their level of confidence in public speaking - http://tinyurl.com/jd3hgcd. After the completion of each prompt, participants were asked about the perceived usefulness of their feedback and difficulty of the prompt. (Control - http://tinyurl.com/zmsvxm4z Treatment - http://tinyurl.com/jxlr7or) In the final survey marking the completion of the study, participants answered questions their experience with the study and system. (Control – http://tinyurl.com/hql5wz6 Treatment – http://tinyurl.com/jxyr2jp)

With self reported surveys, respondents might answer the questions in such a way to please the experimenters. To eliminate this potential bias, we had users fill the surveys anonymously.

# 5. Results

We summarize our analyses of the user study on three key elements: speaking performance, analysis of subjective feedback and system usability.

## 5.1 Speaking Performance

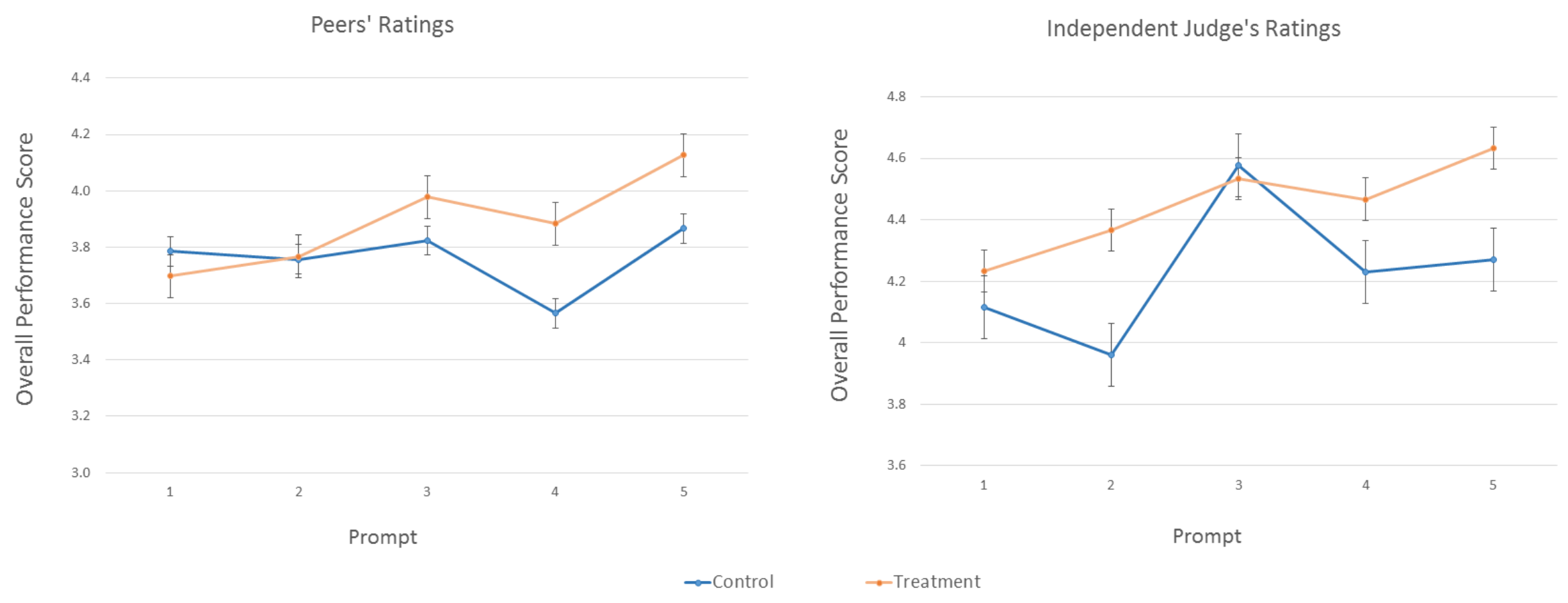

**Figure 4. From left to right, average performance rating by prompt rated by peers (a) and by independent judge (b). Error bars are standard error.**

### 5.1.1 Trajectory of Speaking Performance

We utilized users' assessment of one another's performances and the assessment by an independent judge to evaluate the performance of the speakers. Note that users can submit multiple videos for each prompt—only the final video for each prompt was used for assessment unless otherwise stated.

Each video was rated by an average of 6 (SD=1.486) and 5.5 (SD=1.812) users in the control and treatment group respectively. The number of ratings and feedback were consistent throughout the video. We used Krippendorf's Alpha [48] to measure the agreement between user rating of the videos. Since raters do not rate all the videos, we decided to use this method to account for missing data and multiple raters. We produced a sparse matrix (26x143 for control and 30x167 for treatment by looking at overlapping videos they rated and treating non-overlapping ratings as missing data. We obtained an $\alpha$ between 0.231 to 0.289. The annotations are acquired from 26 and 30 coders over 143 and 167 videos respectively. Because the data is sparse over an ordinal scale, and speech evaluation is an inherently subjective task, it is difficult to compare these $\alpha$ values to the convention acceptable values. Naim et al had a similar task for coding speech performance and acquired $\alpha$ values between

-0.40 to 0.36 over different attributes of speech. They used their annotations to predict interview performance with high success rate [44]. Our $\alpha$ values suggest acceptable agreement between raters in this context.

We averaged the overall performance rating of the users on each video. Figure 4(a) is the trajectory for average ratings for all the users on each prompt. Similarly, Figure 4(b) is created with the same procedure using the ratings of the independent evaluator averaging all user ratings within each prompt. The trajectories indicate a similar pattern of linear increase for user speaking performance with repeated practice. However, the treatment group shows a greater improvement after each session with the exception of the independent judge's data in the 3rd prompt.

*Prompt Difficulty*: The trajectory for performance rating scores for all four trajectories (as shown in Figure 4 (a) and (b)) follow a similar trend. All locally peak at prompt 3, before declining in prompt 4, and (with the exception of the trajectory for independent judge's ratings for control) reaching peak performance in prompt 5. In fact, the trends for the treatment group evaluated by peers and independent judge are nearly identical. These trends can be attributed to the difficulty of the prompt. After the completion of each prompt, we asked the participants about their difficulty. The participants reported average difficulty of the prompt to be in linearly increasing order with a strong correlation (r=0.994). For example, prompt 4 was "describe a challenge or conflict you face," which users reported as the most difficult to answer. However, variability in the patterns such as in prompt 2 maintain that other influences such as individual user's susceptibility to treatment and evaluators' perceptions may play a role as well. A follow up study more specifically investigating this issue could reveal greater insights.

### 5.1.2 Speaking Performance Improvement Assessment

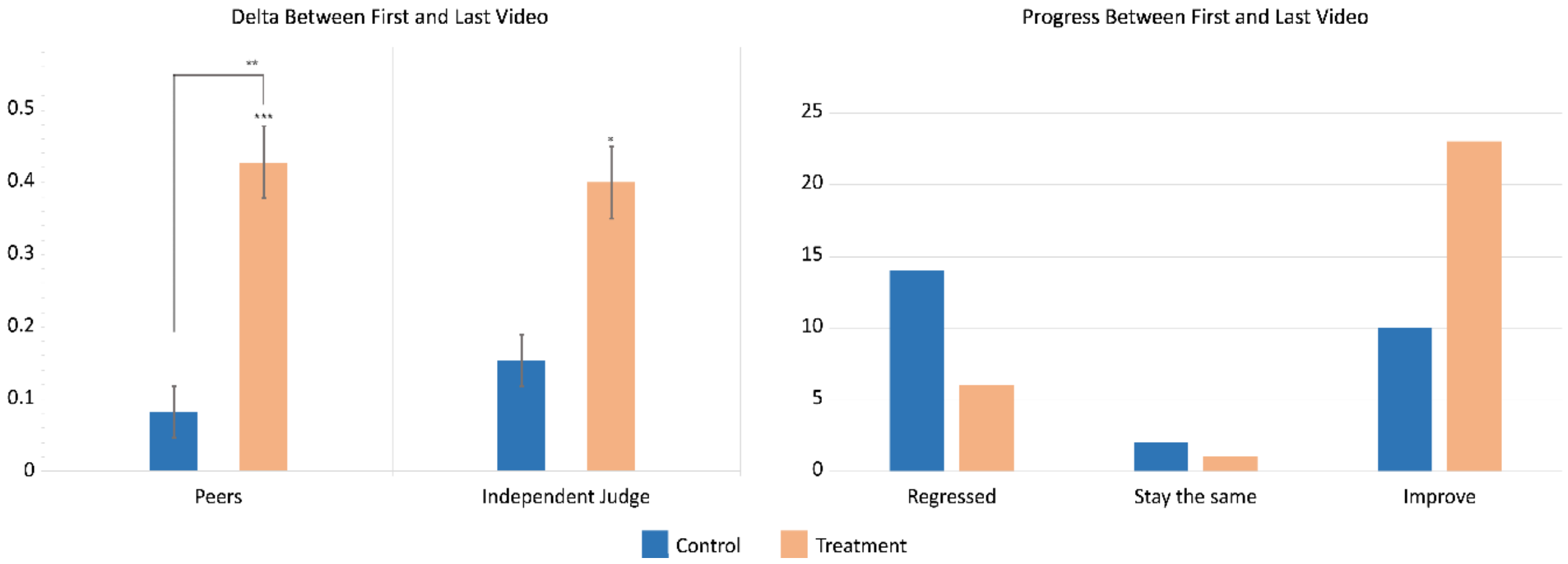

**Figure 5. (a) Performance delta between last and first prompt evaluated by Peers and Independent Judge. (* -> *p*<0.05), (**->*p* < 0.01), (*** -> *p* < 0.001) with t test; (b) Number of people who have a change in ratings for first and last video categorized by a negative (regressed), zero (stayed the same), or positive (improved) delta.**

To validate the effectiveness of our treatment system on user improvement, we studied the user's overall performance scores from the first and last video. These videos addressed the same prompt, thus making them ideal candidates to understand the change in user performance (Figure 5a). The control group, which only had access to subjective feedback from other users, showed a slight improvement in performance. In the user ratings, we observed an average change of 0.09 points (from an average user score of 3.78 to 3.87, n = 26). To analyze the significance of this change, we performed a paired t-test of user performance between these two days and obtained a non-significant result ($p = 0.357$). However, when analyzing changes in the treatment group, we discovered a change of approximately 0.43 points. This highlighted a significant improvement ($p = 2.73\times10^{-5}$, $p < 0.001$) in the groups with access to automated features. Similarly, results from our independent evaluator also suggest that treatment group improves significantly ($p = 0.031$, $p < 0.05$) while the control group does not ($p = 0.42$). To validate a difference in the two groups, we also performed a t-test between their deltas (Figure 5a). We found a difference between the two groups ($p = 0.006$, $p < 0.01$) with peers' ratings but not with the independent evaluator ($p = 0.345$). Although there were improving trends in both conditions, our results suggest that our treatment interface was more effective at improving communication skills than the control interface.

We further analyzed the data by looking at the effect size between the performance improvement of both control and treatment. For each user *u* we measured the difference $D_u = r_i^u - r_f^u$ where $r_i^u$ corresponds to the rating r obtained by user *u* in the initial prompt, whereas $r_f^u$ corresponds to the one on the final prompt. The effect size is reported by Cliff's delta d [49] = 0.468 and Cohen's d [50] = 0.785 (Cohen's d = .8 is attributed to large difference [50]), further suggesting that there is a difference in improvement between control and treatment.

#### 5.1.3 Speaking Attributes Assessment

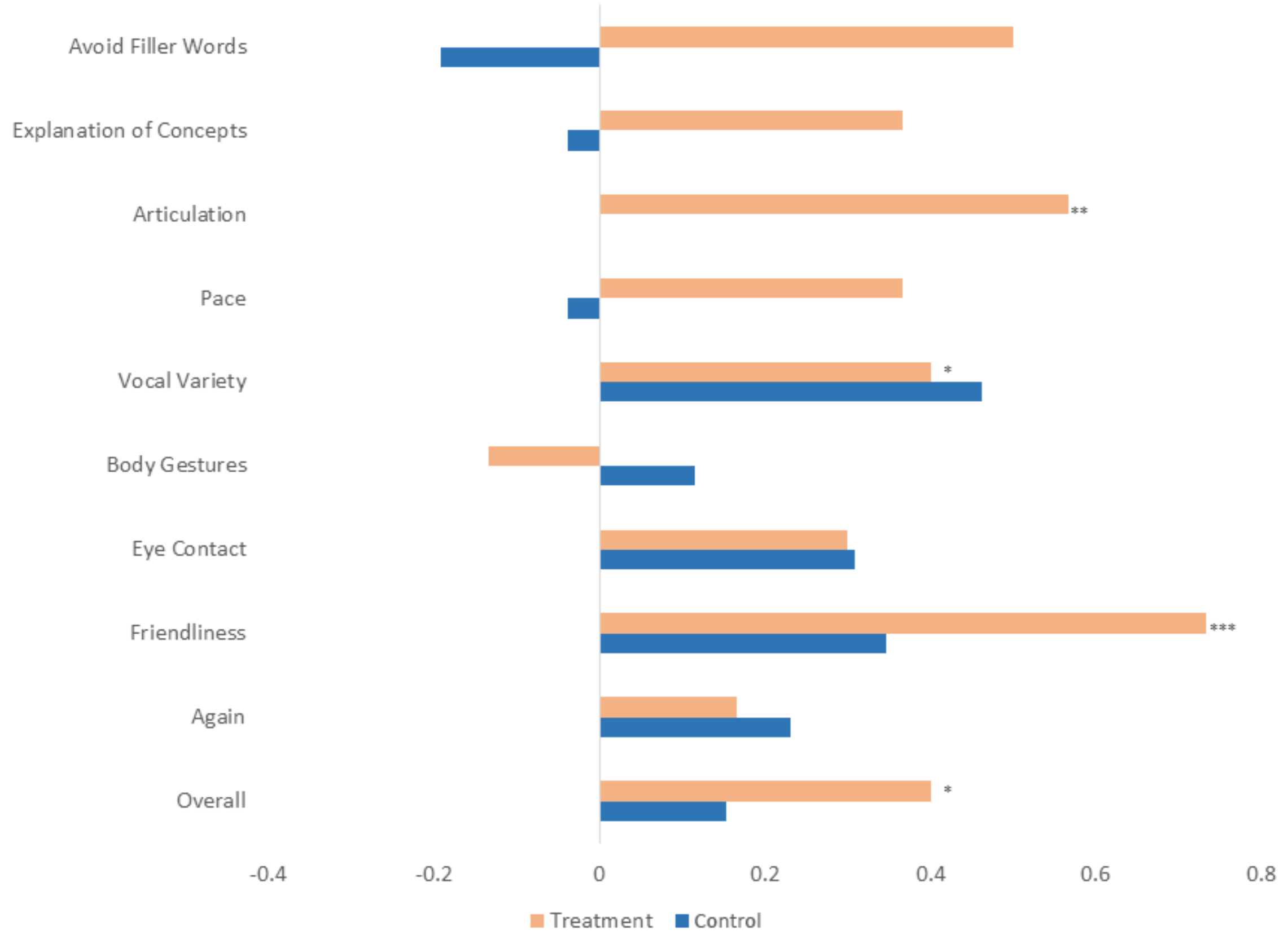

**Figure 6. Post (5th speech)– Pre (1st speech) by Category, (*) denotes p<0.05, (**) denotes p < 0.01, and (***) denotes p<0.001 for test of statistical significance with a t-test.**

Speaking performance can generally be assessed by a number of speech attributes such as articulation and gesturing. To find changes in these characteristics after using our system, we looked at the difference between the first and last video of each user labeled by the independent evaluator. Figure 6 shows the average difference for all users between each prompt for both control and treatment group. Treatment shows a consistently higher change in performance for most attributes. We performed a paired t-test on each of the attributes based on the scores and obtained a statistically significant difference in our delta scores in areas such as Friendliness ($p$ = 0.0006, $p < 0.001$), Vocal Variety ($p$ = 0.049, $p < 0.05$), and Articulation ($p$ = 0.008, $p < 0.001$) in the treatment group while some areas such as Avoid Filler Words, Explanation of concept and Pace were approached statistical significance. No features in the control group displayed any statistically significant improvement in any aspects between the first and last video. Increases in eye contact were consistent between both groups. This was expected because our system does not provide any automated feedback on eye contact. We also saw that body gesture showed a decrease in performance for the treatment group. This can be explained by the form of body gesture measurement we provided. Body gesture is very difficult to measure; our system uses background subtraction to provide some level of movement metric, but it may be too general of an indicator for assessment by the feedback receiver. For example, a proper body gesture would mean having an upright posture, expansion of the shoulder and occasional use of hands. Our future work will involve implementing features such as head nod/ head shake detection, posture, and hand movements identification for more helpful and specific feedback.

In evaluating a speaker's quality of speech, categories such as filler words, eye contact, and friendliness can generally be assessed with confidence, but attributes such as pace, body gestures, and articulation lack ground truth and can heavily depend on an evaluator's own judgment. Furthermore, relying on the expertise of a single rater for these assessments can leave room for error. Nevertheless, the evaluator rated all the videos in random order and was not aware of which group the video came from. The differences found in the two groups carry evidences that our system was effective.

## 5.2 Analysis of Subjective Feedback

### 5.2.1 Overview

In the 10-day study there was no direct intervention from the researchers. The study involved the exchange of 1795 pieces of feedback and the generation of 8916 ratings, 5530 comments, through the interactions of 56 people, fully self-regulated with system constraints.

### 5.2.2 Quality of the Subjective Feedback

Although there was a monetary incentive to leave feedback, the quality of feedback and the qualitative analysis of our survey responses indicate that users received valuable feedback and genuinely wanted to help each other improve. The survey findings below are based on submissions that are anonymous; there were no incentives for them to report positively about the system.

80% (25/31) users in the variable group reported the comments to be helpful, most among all the features. (We did not ask the control group because that's the only type of feedback they see.)

### 5.2.3 Community Development

Despite the feedback being anonymous, users felt a strong sense of community engagement. After completing the study, the treatment group reported a 5.87 (*SD* = 1.76) and the control groups reported a 5.79 (*SD* = 1.41) out of 7 when asked if they felt like they were part of a community. Both groups reported a higher community engagement (p = 0.0001) compare to the expected mean (of 4). The difference (p = 0.866) between the two groups is reported to be insignificant. This suggests that the peer commenting system has a positive effect on community development.

### 5.2.4 Confidence Boost

In an initial survey, participants were asked about their confidence in public speaking. Participants in both the control and treatment group reported an average of 4.16 (control *SD* = 1.73, treatment *SD* = 1.84) out of 7 in the Likert scale. After the study was over, we again asked the users about their confidence in public speaking. The participants reported an average of 5.30 (*SD* = 1.05) and 5.23 (*SD* = 1.41), respectively. Both groups reported to experience an increase in self-confidence (control p = 0.006, treatment p = 0.014) about their public speaking.

### 5.2.5 Progress Tracking in the Comments

Even though the comments were anonymous, there was a trend for some individuals to keep going back to a set of other users. This elicited a sense of progress tracking through the comments. For example, users were able to synthesize information from the previous videos and identify the aspects that the user was able to improve. After investigating peer comments on user videos, we found comments such as:

> “Woah this was a huge improvement from the first introduction”

> “i remember your first video i believe you stated something about your anxiety getting the best of you, and [now you seem] so natural and easy going”

The majority of the users gave more than one feedback to at least one user. This could also be attributed to the participant size, where users have a limited selection of users to provide feedback to. Thus, we speculate that the size of the participant pool can have a direct effect on the level of engagement and speech improvement. To test for such effect, we computed the number of peers each user has interacted with throughout the experiment. Surprisingly enough, the average number of peers users interacted with was 13.9 (SD=5.5) for the control group and 15.5 (SD=6.8) for the variable group, approximately 50% of the total users in each group. Nevertheless, 56% of the users in the control group and 53% in the variable group evaluated at least three different videos of a same user, evidencing that indeed the participants had developed a preference for some of the users and therefore focused their attention in helping them.

As an illustrative example, we observed that a participant (ID 36) provided feedback to a total of 12 videos from a sample of 167 videos. Of these feedbacks, 5 were given to all the videos uploaded by another participant (ID 10). This situation is unlikely to occur by random chance, indicating that participant 36 was “following” participant 10. Our system provides a structure for users to track other users’ progress and provides users with a strong sense of community encouragement.

## 6. **Discussion and Future Work**

The ROC Speak system leverages the strength of human judgment and automated analysis to generate feedback to improve individual’s speaking skills. Our approach is motivated by the recent advancement of augmenting human judgment and computation to perform tasks for clustering complex information [51]. Such hybrid systems leverage the combined strengths of humans and computers and can outperform either alone. We believe that the ROC Speak system encompasses some aspects of that strength by generating meaningful feedback that is more powerful than feedback generated from a machine or a human alone. For example, our automated feedback provides unbiased raw data on a user's smile intensity. Without a large dataset and annotations, we cannot accurately suggest to a user if he/she should smile at a given point. However, humans can intuitively judge if a smile is appropriate based on prior experiences in communicating with others. At the same time, people lack the ability to track the exact moment they smiled and their judgment is often influenced by what they remember, as evidenced by their capacity to remember only the first and last items of a list [52]. As a result, human judgement alone cannot evaluate all aspects of a speech. Carefully looking at features such as smiling or body movement would require a peer to stop the video frequently and can be very expensive to annotate. Given the improvements of our treatment group over the control group in our study, we believe that there is potential in systems that leverage the strength of both types of feedback.

Results from our 10-day intervention showed that individuals who interacted with our combined automated and subjective feedback system could significantly improve their speaking skills more than a control group. However, while these improvements can be attributed to ROC Speak’s feedback system it is unclear which components, the automated feedback or the peer feedback organizer, are contributing to these improvements. Our next step is to run another experiment that decouples the effects of the automated feedback and peer feedback organizer. This may let us gain new knowledge on feedback generation that may be helpful beyond the scope of communications skills training.

The quality of assessments in some attributes of speech is heavily dependent on the rater as a result of lack of ground truth. We can improve the quality of these assessments by providing the rater a rubric and initial training. Furthermore, including multiple evaluators will reduce the subjectivity in these assessments.

Our intervention included five exposures to the system over a 10-day period. While the duration of the study is an improvement over the state-of-the-art, a long term study may reveal more insights. An effect over a 10-day period could be a priming effect as opposed to a permanent change in speaking delivery. In our future work, we intend to run a longer study to test whether the changes that result from using the ROC Speak system is permanent or not.

In our study, we recruited online workers from Amazon Mechanical Turk (AMT) for feedback generation and as participants. While this method is convenient and offers a diverse set of participants [53, 54], there are concerns about the quality of the data collected and worker motivation. Previous work in natural language tasks has shown that gathering responses from numerous non-expert workers can lead to results comparable to expert annotators [55]. Other works have shown that crowdsourcing can be used in various applications such as video annotations [56], editing written work [57], and real-time captions [58] and produce quality results. These studies have motivated our use of Turkers as a source of data for our trained classifiers and as participants who provide interpretive feedback on videos.

As Turkers left comments for each other, they were aware of the fact the experimenters could review their comments. This could have had a bias in the way they expressed their views on others videos. While the bias was consistent across the two groups, whether the participants would express their view differently, if not watched, remains for further investigation.

A major privacy concern with the design of the ROC Speak system was that individuals would have to opt in to share their videos with others. Participants may prefer to receive personalized and subjective feedback yet be hesitant to share their videos with others. Can an automated system automatically generate positive and negative comments given a behavioral video? With the advancement of automated image captioning methods, this idea does not seem far-fetched. One advantage of ROC Speak is that it allows researchers to collect a large corpus of labeled and commented videos which can be used as a trained dataset. In our initial dataset of 196 public speaking videos from 49 individuals, we were able to gather 12,173 comments from 500 independent human judges. Then, we trained a k-nearest neighbor (k-NN) that can, given a new video, select appropriate set of comments [59]. Incorporating this feature into ROC Speak would be an exciting experimental possibility. With this feature, we could see if there is a performance difference for people whose comments are generated by humans or by a computer algorithm.

During the study. we found that some participants tended to go back to the same users to provide feedback perhaps because they enjoyed watching that user or wanted to see the user's progress. This phenomenon could be evaluated as a social network where individuals are treated as nodes and exchanged feedback as edges. Analyzing these interactions could reveal interesting insights about how users form sub-communities and the varying dynamics of these communities. For instance, all the good speakers could form a community, all the bad speakers could form a community, or a mixture of good and bad speakers could form a community. Upon learning which type of communities work best, we could artificially manipulate the way people exchange feedback by accounting for the ratings they received in earlier prompts. For example, we could adjust the video feed page such that users who received positive ratings in their earlier videos would see users that had previously received bad ratings at the top of their feed. Through this manipulation, users that initially received bad ratings would receive feedback from users that had already demonstrated excellence in their previous speech. This could add another aspect of automation to our system in feedback generation. Studying these interactions is part of an ongoing work and may appear in a future publication.

In a survey in 2012 [60], it was argued that more than 60% of screening interviews are now being conducted using videos. With the advancement of videoconferencing tools, this number is expected to grow. There are no tools available that provides individuals with ubiquitous access to practice and receive standardized and respectful feedback. The framework of ROC Speak and its validation could potentially fulfill this need while allowing us to collect a large spontaneous dataset. Although we validated our framework in the context of interviews, it can be customized to enable other types of communication skills training such as public speaking, tutoring, and dating.

With companies receiving lots of job applications, some begin to opt for pre-recorded videos as part of their initial screening process [61, 62]. These video interviews don't require scheduling with an interviewer, allowing recruiters to reach more applicants [63]. There are now companies such as Sonru [64] and HireVue [65] that offer the technology and professional services to enable this process. Even many colleges now include a video essay as part of their application package [66, 67]. It is important to develop new tools that could help interviewees to keep up with this new video interview process. The ROC Speak framework and its extensive evaluation presented in this paper could be useful for the general public to prepare for their interview process. Users can upload a video to our system and use the feedback to get a better sense of how the recruiters may perceive their performance. Furthermore, if the users opt in to share their data, it could inspire new machine learning techniques. For example, ROC Speak could inform users with a prediction of success for their application from their videos.

## 7. Conclusion

In this paper, we present a training module, called ROC Speak, to aid speaking skills for participants whenever they want, wherever they want using a computer browser. We recruited 56 online workers and demonstrated that the participants who received the feedback improved significantly more than a control group. Our study was conducted in a real-world setting with minimal intervention, eliminating the effects of confounding factors that may come from lab settings, interactions with researchers and novelty factors. This paper also addressed the potential application and limitations of the ROC Speak system. The entire framework is readily available for deployment and ubiquitously available for anyone with access to a browser and a webcam.